\documentclass[a4paper,fleqn]{cas-sc}
\usepackage[numbers,sort&compress]{natbib}
\usepackage{lineno}
\usepackage[section]{placeins}
\usepackage{float}

\def\tsc#1{\csdef{#1}{\textsc{\lowercase{#1}}\xspace}}
\tsc{PS}
\tsc{fs}

\newif\ifPRODUCTIONFLOATS \PRODUCTIONFLOATSfalse

\begin{document}

\renewcommand{\topfraction}{0.92}
\renewcommand{\bottomfraction}{0.85}
\renewcommand{\textfraction}{0.06}
\renewcommand{\floatpagefraction}{0.80}
\setcounter{topnumber}{3}
\setcounter{bottomnumber}{2}
\setcounter{totalnumber}{4}

\let\WriteBookmarks\relax

\shorttitle{fs-Laser Suppression of Microsphere Sedimentation}

\shortauthors{Moazzamia et~al.}  

\title [mode = title]{Suppression of Sedimentation in Polystyrene Microsphere Suspensions via Femtosecond Laser Irradiation}  

\author[1]{Hafsa Moazzamia}[orcid=0009-0004-2150-3923]
\affiliation[1]{organization={Department of Physics, Middle East Technical University},
            city={Ankara},
            postcode={06800}, 
            country={Turkey}}
\author[1]{Irem Hazal Tan}[orcid=0009-0006-8796-6984]
\author[1,2]{Alpan Bek}[orcid=0000-0002-0190-7945]
\author[1,3]{Sinan Genc}[orcid=0000-0002-6909-723X]
\cormark[1]
\ead{sinan.genc@agu.edu.tr}
\affiliation[2]{organization={Department of Micro and Nanotechnology, Middle East Technical University},
            city={Ankara},
            postcode={06800}, 
            country={Turkey}}
\affiliation[3]{organization={Department of Electrical and Electronics Engineering, Abdullah Gül University},
            city={Kayseri},
            postcode={38080}, 
            country={Turkey}}
\cortext[1]{Corresponding author}

\begin{abstract}
Micrometer-sized particles denser than their medium settle out of suspension within hours, and Brownian motion is too weak to prevent it. We report that femtosecond laser irradiation at 1030~nm slows this settling in suspensions of 8~$\mu$m polystyrene microspheres in water, without altering the particles themselves. Suspensions at three concentrations were treated for up to 3~h under weak acoustic agitation or vigorous magnetic stirring, each with and without simultaneous irradiation, and the particles remaining in the probe volume were quantified optically by angle-resolved Mie scattering. Under acoustic agitation, irradiation raised retention in eleven of twelve paired comparisons, at every concentration, by 0.17 on average against a measured aliquot-to-aliquot scatter of 0.02; the twelfth comparison lies within that scatter. Under stirring, which already keeps most particles suspended, it produces no resolvable change. The angular positions of the Mie maxima are unchanged to within about 1\% over 3~h, so the particles are neither fragmented nor aggregated and the effect arises from particle transport rather than from changes in the particles. An order-of-magnitude analysis rules out optical forces, the optical Kerr effect, and transient photochemistry, and leaves laser-driven convection: the power absorbed by the water sustains flows orders of magnitude faster than the 1.8~$\mu$m/s settling velocity. Femtosecond irradiation therefore acts as a non-contact, supplementary agitation source, most effective where mechanical mixing is weak.
\end{abstract}


\maketitle

%% ---------------------------------------------------------------- 1 INTRODUCTION
\section{Introduction}\label{sec1}

Suspension stability is a critical parameter across a broad range of scientific and industrial domains, including pharmaceuticals, food science, cosmetics, and materials engineering. For sufficiently small particles, Brownian motion counteracts gravitational settling and sustains a stable colloid without external intervention. This protection vanishes rapidly as particle size increases. For a polystyrene (PS) sphere of 8~$\mu$m diameter in water (density contrast $\Delta\rho \approx 0.05~\text{g/cm}^3$), the buoyancy-corrected weight is only ${\sim}0.13$~pN, yet the corresponding gravitational length $l_g = k_BT/(\Delta m\, g) \approx 30$~nm is more than two orders of magnitude smaller than the particle radius, and the Stokes settling velocity is ${\sim}2~\mu$m/s, or ${\sim}7$~mm per hour \citep{Guazzelli2012}. Such particles therefore leave an optical probe volume within tens of minutes to hours, and their suspensions cannot be stabilized passively. Particles in this size range are ubiquitous: they include human blood cells \citep{Kinnunen2011,Tigner2025}, many environmental microorganisms \citep{Felter1969,Fradette2022}, and microplastic particles \citep{Genc2023,Genc2024}.

Established strategies to stabilize suspensions act on particle--particle and particle--medium interactions. pH modulation alters the surface charge of dispersed particles, tuning electrostatic repulsion and thereby aggregation and settling \citep{Xiong2021,Huang2009,Ife2018}. Steric stabilization through adsorbed or grafted polymers provides a complementary route, although its effectiveness depends on particle morphology and environmental conditions and long-term stability is limited by polymer desorption \citep{Rawlins1980,Grelet2016,Wang2020}. Thermo-responsive coatings exploit polymer solubility transitions to modulate inter-particle interactions \citep{Pavlovic2020}, and composite approaches combine surface functionalization with nanofluid engineering. These methods require chemical modification of the particle surface or the medium, which is undesirable in biological and environmental samples, and, more fundamentally, they suppress aggregation rather than gravitational settling of individual particles. For non-Brownian particles, sedimentation must instead be counteracted by fluid motion. In practice, this is supplied by mechanical agitation such as magnetic stirring or acoustic excitation, whose effectiveness depends on the flow field it generates \citep{Wiklund2012} and which requires either contact with the sample or mechanical coupling to the container.

Lasers offer a non-contact alternative. Laser processing of colloids is a mature field, in which pulsed irradiation is used to synthesize, fragment, and modify particles in liquids \citep{Zhang2017}; fs ablation in liquids has, for example, converted sedimenting dispersions of large nanoparticles into stable transparent colloids by reducing particle size \citep{Dai2017}. Seo~et~al.\ reported substantially enhanced stability of aqueous alumina nanoparticle suspensions after fs irradiation at 800~nm, attributed to fragmentation of particle clusters and modification of surface chemistry, with stability maintained for up to 30 days \citep{Seo2015}. Femtosecond irradiation has similarly been shown to modify aqueous TiO$_2$ nanoparticle suspensions through stabilization at low fluence and fragmentation at high fluence \citep{HA2019662}. All of these studies, however, concern nanoparticles in the Brownian regime, where stability is governed by particle size, surface charge, and cluster size, and they achieve stabilization by permanently changing the particles. Neither mechanism can counteract the settling of individual micrometer-sized particles whose size is to be preserved, and whose stability is set by the balance between gravity and fluid motion.

Light can, however, act on this balance directly, both on the particles and on the fluid. Optical gradient and scattering forces trap and propel dielectric microparticles \citep{Ashkin1970,Grier2003}, including under fs excitation \citep{Agate2004,Xing2004}, and radiation pressure combined with photothermal convection has been used to transport 5--50~$\mu$m particles at tens of $\mu$m/s over sub-millimeter distances \citep{Li2025}; given the sub-piconewton buoyant weight estimated above, optical forces cannot be dismissed \emph{a priori} for particles in the beam path. Absorption of near-infrared light by water, weak but non-negligible at 1030~nm \citep{HaleQuerry1973,Kou1993}, creates temperature gradients that drive buoyancy-driven convection and thermophoresis \citep{BraunLibchaber2002,DuhrBraun2006,Riviere2016}, and such optothermal flows have been used to transport, trap, and assemble microparticles \citep{LiuPoon2010,RamirezRamirez2023}. At the intensities reached by focused fs pulses, nonlinear absorption in water additionally generates free electrons, transient plasma, pressure transients, and cavitation or long-lived gas bubbles \citep{Vogel2005,Vogel2008,Couairon2007,Sabzeghabae2021}, whose oscillation and motion induce secondary flows \citep{Lauterborn2010}. Each of these processes can, in principle, generate fluid velocities exceeding the ${\sim}2~\mu$m/s settling velocity of 8~$\mu$m PS spheres. They have, however, been studied predominantly in thin microfluidic layers and for the manipulation of individual particles. Whether fs irradiation can counteract sedimentation of a bulk suspension in a macroscopic cell, and which of these mechanisms dominates, has not been established.

Answering this question also requires a stability readout that distinguishes the loss of particles from the probe volume from changes of the particles themselves. Conventional turbidity and multiple-light-scattering analyzers track transmitted or backscattered intensity \citep{Mengual1999}, which decreases both when particles settle and when they aggregate or fragment. Angle-resolved Mie scattering resolves this ambiguity: the angular positions of the scattering maxima are set by particle size and refractive index, while the overall scattered intensity increases with the number of particles in the illuminated volume \citep{Hulst1981,Genc2023,Genc2024,Genc2024plp}.

In this work, we show that fs irradiation at 1030~nm retards the sedimentation of 8~$\mu$m PS microspheres in deionized water and assess the origin of the effect. We make four contributions. First, we report the complete irradiation, acoustic, stirring, and detection parameters, together with the measurement sequence, so that the experiment can be reproduced. Second, we introduce an integrated retention metric, $R(t)$, computed over the full accessible angular range of the Mie profile; we show that it is a monotonic and conservative measure of the particle number in the probe volume, that it is insensitive to the integration limits, and we use the invariance of the Mie peak positions to show that particle size is preserved. Third, we compare four processing conditions, acoustic agitation or magnetic stirring each with and without simultaneous irradiation, at three particle concentrations over 3~h, and find that irradiation substantially increases retention when mechanical mixing is weak and has no resolvable effect when it is strong. Fourth, we evaluate the candidate mechanisms quantitatively against the Stokes settling velocity and show that laser-driven convection in the medium is the only mechanism of sufficient magnitude. Together, these results identify fs irradiation as a non-contact supplementary agitation source for non-Brownian suspensions.

%% ---------------------------------------------------------------- 2 METHODS
\section{Materials and Methods}\label{sec2}

\subsection{Particles and sample preparation}\label{sec2.1}

Monodisperse PS microspheres (PS8; Supelco/Sigma-Aldrich, product no.\ 84192, lot BCCM8580) were used as received. According to the certificate of analysis, the particles have a calibrated diameter of $7.989 \pm 0.097~\mu$m (coefficient of variation 1.2\%, Coulter Multisizer), a density of 1.05~g/cm$^3$, a solid content of 2~wt\%, and a stock concentration of approximately $7.1\times10^{7}$ particles/mL (${\approx}118$~fM). Brightfield optical microscopy of the stock confirms the morphology: the particles are discrete, smooth spheres with no fused or irregular objects and no aggregates (Supplementary Fig.~S1). Edge detection on 60 in-focus singlets in that field gives a diameter of $7.78 \pm 0.19~\mu$m (coefficient of variation 2.4\%), consistent with the certificate; the small offset and the larger spread are set by the diffraction-limited edge of a brightfield image and by residual defocus, so the certified value is used throughout. The measured axial ratio is $1.01 \pm 0.02$. The stock was vortexed for 1~min at 500~rpm before each dilution. Suspensions of 1, 2, and 3~fM, corresponding to $0.60$, $1.20$, and $1.81\times10^{6}$ particles/mL and particle volume fractions of $1.6$--$4.8\times10^{-4}$, were prepared by dilution in commercial bottled deionized water using micropipettes with disposable polypropylene tips. At these volume fractions, the mean interparticle distance (${\approx}80$--$120~\mu$m) exceeds the particle diameter by an order of magnitude, so that hydrodynamic interactions between particles are weak and particles settle essentially independently. Experiments were carried out at ambient laboratory temperature over several months; the room temperature was not controlled.

\subsection{Sample cell}\label{sec2.2}

Each sample (1.1~mL) was placed in a disposable PS macro cuvette with clear walls (ISOLAB, 098.02.002; external dimensions $12.5 \times 12.5 \times 45$~mm, internal cross section $10 \times 10$~mm, 10~mm optical path), which has two transparent optical faces and two slightly tapered, non-optical side walls. At the filling volume used, the liquid column height is ${\approx}11$~mm. Although the cuvette is specified for spectrophotometry between 340 and 900~nm, PS is transparent at 1030~nm \citep{Wang2019}, and the fs beam was focused inside the liquid, so that the intensity at the cuvette wall is far below that at the focus. No damage of the cuvette walls was observed after irradiation. The same cuvette type and filling volume were used for all treatments and measurements, and cuvettes were sealed with Parafilm during processing to prevent evaporation.

The geometry of the cell and of the two beams is shown in Fig.~\ref{fig1}. Panel (a) is a side view, giving the 11~mm liquid column, the height of the fs and probe beams above the cuvette bottom, the focal position at the center of the cell, and the magnetic stir bar. Panel (b) is a top view, showing the converging fs beam entering through one transparent face with its focus 5~mm behind the entrance wall, and the two earphone drivers coupled to the tapered side walls. Panel (c) is the measurement sequence: each treatment time is a separate aliquot of the same concentration, so that the four points of a series are independent samples rather than successive observations of one sample.

\begin{figure*}[t]
  \centering
  \includegraphics[width=0.82\textwidth]{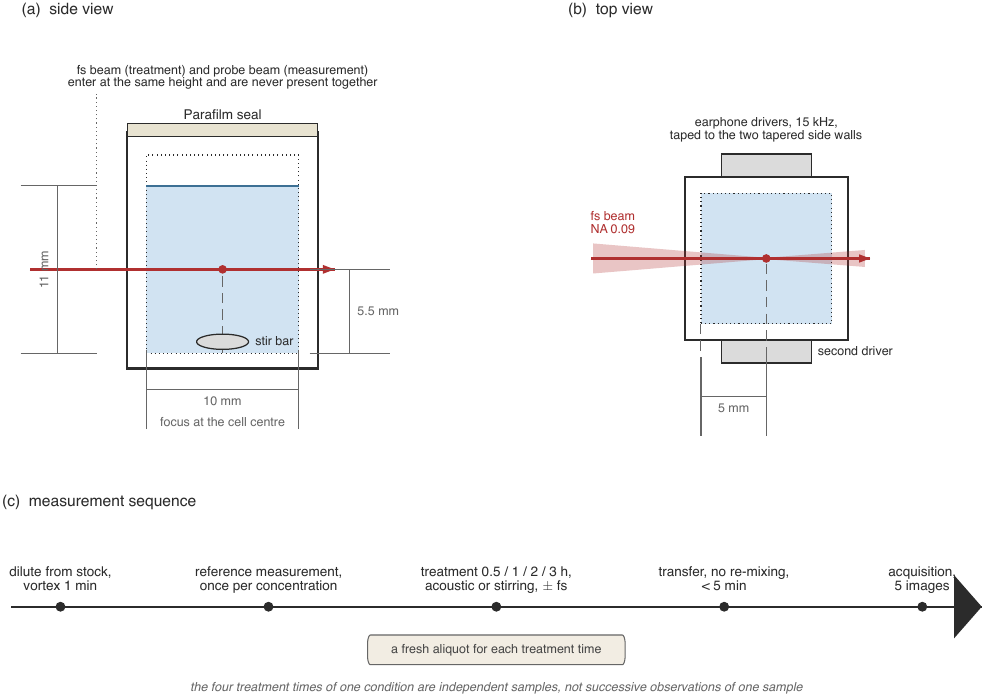}
  \caption{Schematic of the experimental configuration: (a) side view of the cuvette, (b) top view of the beam and acoustic geometry, and (c) the measurement sequence. The fs beam is present only during treatment and the probe beam only during measurement, and the two are never present together.}
\end{figure*}

\subsection{Femtosecond irradiation}\label{sec2.3}

Irradiation was performed with an in-house-built mode-locked fiber laser system \citep{Erkizan2022} delivering pulses at a central wavelength of 1030~nm with a duration of 370~fs, measured by autocorrelation at the laser output, at a repetition rate of 1~MHz. Dispersion in the objective and the 1~mm PS entrance wall (group delay dispersion of the order of $10^{2}$--$10^{3}$~fs$^2$) is negligible compared with $\tau^2$ for 370~fs pulses, so the pulse duration at the focus is essentially that measured at the output. 

The average power was set with a half-wave plate--polarizing beam splitter--half-wave plate assembly, which allows continuous attenuation without changing the beam path. The laser provides up to 2\,W; the average power was set with the attenuator and measured with a power meter at the sample position to be 1.8\,W, corresponding to a pulse energy of 1.8\,$\mu$J. The linearly polarized beam was focused through one transparent face of the cuvette by the laser objective (working distance 9~mm in air, specified focal spot radius 2--5~$\mu$m). For a beam waist of $w_0=3.5~\mu$m, the beam radius at the objective is 0.84~mm and the far-field half-angle is 0.094~rad in air (NA $\approx0.09$) and ${\approx}4^\circ$ in water. Refraction at the air--cuvette and cuvette--water interfaces, calculated from Snell's law and Gaussian beam propagation, displaces the focus to 10.63~mm from the objective; the objective was positioned 4.63~mm from the outer surface of the entrance wall, such that the focus lay at the center of the cell, 5~mm behind the inner surface of the entrance wall (objective--cuvette distances of 3.12--6.13~mm, corresponding to focal depths of 3--7~mm, were taken as the tolerance) and at half the liquid height (${\approx}5.5$~mm above the cuvette bottom). Because the objective remains in air, the waist in water is unchanged ($w_0 \approx 3.5~\mu$m). The beam was stationary during irradiation. The resulting peak fluence is $F_0 = 2E_p/(\pi w_0^2) \approx 9.4$~J/cm$^2$ and the peak intensity $I_0 = 0.94\,F_0/\tau \approx 2.4\times10^{13}$~W/cm$^2$, above the threshold for fs optical breakdown in water \citep{Vogel2005,Vogel2008}. The Rayleigh length in water is ${\approx}50~\mu$m, so that the high-intensity focal region occupies only ${\sim}10^{-9}$ of the sample volume. All parameters are summarized in Table~\ref{tbl:laser}.

\begin{table}[pos=ht]
\caption{Femtosecond irradiation parameters.}\label{tbl:laser}
\centering\small
\setlength{\tabcolsep}{4pt}
\renewcommand{\arraystretch}{1.15}
\begin{tabular}{@{}p{0.26\textwidth}p{0.40\textwidth}p{0.28\textwidth}@{}}
\toprule
Parameter & Value & Determination \\
\midrule
Central wavelength & 1030~nm & Laser system specification \cite{Erkizan2022} \\
Pulse duration & 370~fs & Autocorrelation at the laser output \\
Repetition rate & 1~MHz & Laser system specification \cite{Erkizan2022} \\
Average power at sample & 1.8~W & Power meter at the sample position \\
Pulse energy at sample & 1.8~$\mu$J & Average power / repetition rate \\
Beam diameter at objective & ${\approx}1.7$~mm & Gaussian propagation from $w_0$ \\
Focusing optics & Objective, 9~mm working distance, NA ${\approx}0.09$ & Manufacturer specification \\
Focal spot diameter, $2w_0$ & ${\approx}7~\mu$m (specified 4--10~$\mu$m) & Manufacturer specification \\
Peak fluence & ${\approx}9.4$~J\,cm$^{-2}$ & $F_0 = 2E_p/(\pi w_0^2)$ \\
Peak intensity & ${\approx}2.4\times10^{13}$~W\,cm$^{-2}$ & $I_0 = 0.94\,F_0/\tau$ \\
Polarization & Linear & HWP--PBS--HWP assembly \\
Focal depth in the cell & $5\pm2$~mm behind the entrance wall & Snell's law, Gaussian optics \\
Focal height & ${\approx}5.5$~mm above the cuvette bottom & Half the liquid column \\
Beam motion & Stationary & --- \\
Irradiation times & 0.5, 1, 2, 3~h & --- \\
Number of pulses & $1.8\times10^{9}$ to $1.08\times10^{10}$ & Repetition rate $\times$ time \\
\bottomrule
\end{tabular}
\end{table}

Two observations made during irradiation are relevant later. Bubbles appeared in the samples after some time of irradiation rather than immediately; they were distributed throughout the liquid and on the cuvette walls rather than only at the focus, remained largely at their positions during exposure, and became more numerous with longer exposure. After irradiation, the samples were noticeably warm, and no deposit or damage was found on the cuvette walls.

\subsection{Acoustic agitation}\label{sec2.4}

Acoustic agitation was applied with a pair of miniature electrodynamic transducers (earphone drivers) pressed against the two tapered side walls of the cuvette (Fig.~\ref{fig2}a) and driven continuously by a 15~kHz sinusoidal signal at the maximum output of the driving circuit (a commercial hands-free earphone set); the drivers were mounted on the translation stage and attached to the cuvette walls with double-sided tape. The electrical drive level was not measured. Acoustic energy was thus coupled mechanically through the cuvette walls. At 15~kHz, the acoustic wavelength in water (${\approx}10$~cm) is much larger than the cell, so no standing-wave pattern forms within the sample; agitation arises from wall vibration and the associated bulk flow. The acoustic pressure in the sample was not measured; the acoustic condition is therefore specified by the source, its drive, and the coupling geometry. Compared with magnetic stirring, this provides weak and spatially inhomogeneous mixing. Its ability to delay settling was checked in a preliminary test on a single freshly prepared batch, measured three times with identical camera settings (Supplementary Fig.~S4). The integrated scattering fell to $0.96$ of its initial value after the sample had stood undisturbed for 10~min, and to $0.99$, that is, unchanged within the frame-to-frame scatter of the measurement, after a vortexed aliquot had been held under 15~kHz excitation for the same interval. The difference is small in absolute terms because 10~min is short compared with the ${\approx}50$~min a particle needs to fall from the liquid surface to the probe height (Section~\ref{sec3.1}); the same measurement therefore also bounds the effect of the transfer delay of Section~\ref{sec2.6}.

\begin{figure*}[t]
  \centering
  \includegraphics[width=0.7\textwidth]{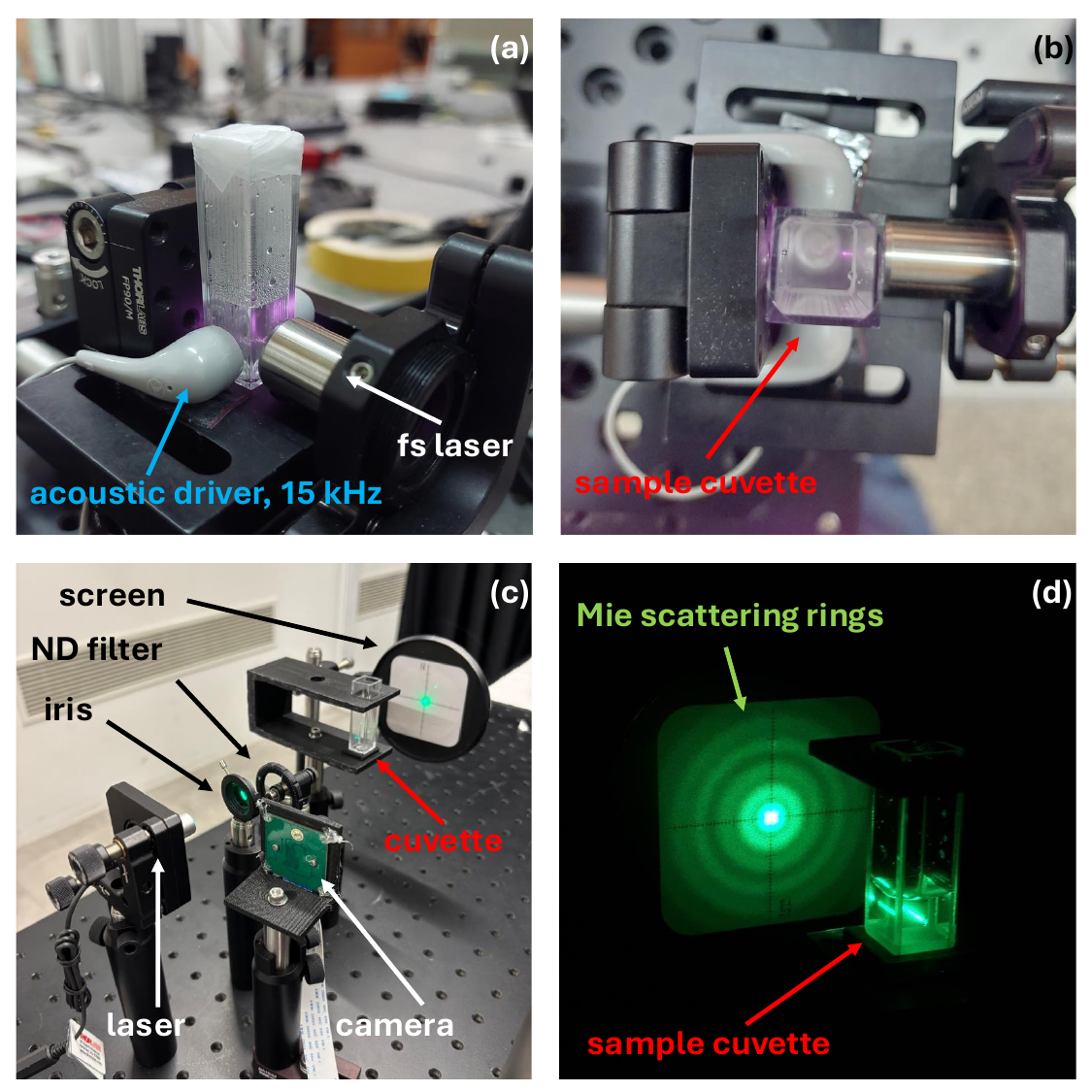}
  \caption{Photographs of the two setups. (a) Femtosecond irradiation assembly: the beam is focused through one transparent face of the PS cuvette, and an earphone driver is coupled to a tapered side wall to provide the 15~kHz acoustic agitation applied simultaneously with irradiation. (b) Close-up of the focusing geometry. (c) Angle-resolved scattering setup, to which treated samples are transferred for measurement. (d) A scattering pattern recorded in the dark in setup (c).}
\end{figure*}

\subsection{Magnetic stirring}\label{sec2.5}

Magnetic stirring was performed on a magnetic stirrer at 500~rpm using a PTFE-coated magnetic stir bar (${\approx}5\times2$~mm) resting on the cuvette bottom. Stirring was applied continuously during treatment and stopped at the end of treatment. The stir bar remained in the cuvette during the scattering measurement; it lies below the probe beam and does not intersect it.

\subsection{Processing conditions and measurement sequence}\label{sec2.6}

Four processing conditions were evaluated at each concentration: (a) acoustic agitation alone, (b) acoustic agitation with simultaneous fs irradiation, (c) magnetic stirring alone, and (d) magnetic stirring with simultaneous fs irradiation. In (b) and (d), irradiation and mechanical agitation were started and stopped together. For each condition, separate aliquots of the same concentration were treated for 0.5, 1, 2, and 3~h; each data point therefore represents an independent sample rather than a repeated measurement of the same sample. The design is thus parallel rather than longitudinal: a given treatment time is not a later observation of the preceding one, and the four points of a series are statistically independent. Unagitated controls were not included because the scattering signal decayed too rapidly for reliable quantification.

Irradiation and scattering measurements were not performed simultaneously. After treatment, the sealed cuvette was transferred to the scattering setup (Fig.~\ref{fig2}c) without any re-mixing, i.e., without vortexing, shaking, or pipetting, and measured within 5~min after the end of treatment. During this interval, particles settle by ${\lesssim}0.6$~mm, much less than the 5.5~mm between the liquid surface and the probe beam, so that settling during transfer does not change the particle concentration at the probe height; the 10~min test of Section~\ref{sec2.4}, in which the integrated scattering fell by only 4\%, confirms this directly. The measurement therefore records the particles that remained in the probe volume at the end of treatment. The fs laser was thus never present during scattering acquisition, and no optical interference between the pump and probe beams could occur; the pattern in Fig.~\ref{fig2}d was recorded in the scattering setup. 

For each concentration, one reference measurement was recorded on a freshly prepared and vortexed suspension immediately after preparation, without any treatment; the treated aliquots themselves were never re-mixed. The four processing conditions at a given concentration were not all run on the same day. To make them directly comparable, all four conditions of a concentration were normalized to this same reference measurement. Because the reference enters every $R$ of that concentration as a common divisor, an error in it rescales the irradiated--non-irradiated difference $\Delta R$ by a common factor but cannot change its sign or the ordering of conditions. The complete sequence is shown in Fig.~\ref{fig1}c.

\subsection{Angle-resolved Mie scattering measurements}\label{sec2.7}

Angle-resolved scattering was measured in a custom-built setup described previously \citep{Genc2024} (Fig.~\ref{fig2}c). A linearly polarized green diode laser (Thorlabs CPS520; the part designation refers to a nominal 520~nm, whereas the manufacturer’s data sheet specifies an emission wavelength of 514.9~nm, which is the value used here) with an optical power of 300~$\mu$W, measured before each acquisition, passed through an iris and a neutral density filter and traversed the 10~mm optical path of the cuvette at half the liquid height (${\approx}5.5$~mm above the cuvette bottom), i.e., at the same height as the fs focus during treatment. Light scattered in the forward direction was projected onto a screen 6.5~cm behind the cuvette and recorded with a Raspberry Pi Camera (5~MP OV5647 sensor) at ISO~100 and an exposure time of 0.1~s. Five images were recorded per sample, about 6.5~s apart. The analysis uses the green channel of the camera, which carries the 515~nm probe light; the red and blue channels are dominated by the automatic white balance of the camera, whose gains change from frame to frame.

The center of the scattering pattern was taken as the centroid of the saturated pixels of the transmitted beam (green value $\geq 250$) in each frame. Owing to the circular symmetry of the pattern, one quadrant was used: the intensity was sampled along 81 radial lines with azimuthal angles from 90$^\circ$ to 170$^\circ$ in 1$^\circ$ steps and averaged over azimuth. Along each radial line, samples beyond the edge of the screen, where the signal falls to the black level of the image, were excluded, so that the azimuthal average contains only points on the screen. The profiles of the five frames were then averaged. The radial distance $r$ on the screen was converted to the scattering angle inside the sample as $\theta = \arcsin[\sin(\arctan(r/L))/n_w]$, where $L$ is the effective distance from the probe volume to the screen and $n_w=1.335$; the refraction correction at the cuvette--air interface amounts to ${\approx}25\%$ at the largest angles used and cannot be omitted. Rather than relying on the nominal value of $L$, the angular axis of every profile was calibrated against the computed Mie pattern of the particles: $L$ was obtained by maximizing the correlation between the measured and the computed logarithmic profiles over 3.5--17$^\circ$, after removal of the slowly varying envelope from both. This uses all Mie orders in that range rather than individual maxima, converges for every measurement, including the most depleted samples, and is applied identically to all of them. Converted to millimeters with the graduations of a ruler on the screen (25.8~px/mm), the fitted $L$ averages 65.2~mm, consistent with the screen distance, and scatters by 0.6\% within a series (Supplementary Table~S4).

Because the cell is removed and replaced between measurements, the distance $D$ between the probe volume and the screen is not exactly the same in every measurement; in one series (2\,fM, acoustic agitation) the fitted $L$ is about 8\% larger than in the others, while the camera and the screen remained in place, as the unchanged pixel positions of the screen edges show. At a given scattering angle, the irradiance on the screen scales as $1/D^2$, and the fitted $L$ is proportional to $D$. Each integrated profile is therefore multiplied by $L^2$ (Eq.~\ref{eq:retention}). One series, 1\,fM under acoustic agitation, was recorded after the scattering setup had been reassembled, with the camera and the screen in a different position from all other measurements. Its normalization to the common 1\,fM reference therefore also spans a change of the detection geometry, which the $L^2$ factor does not correct. An untreated 1\,fM suspension measured in the reassembled setup scatters 0.63 times as much as the common reference; normalized to that measurement instead, three of the four 1\,fM acoustic comparisons change sign (Supplementary Table~S5).

One frame of the 0.5~h acquisition of the 1\,fM series under acoustic agitation was discarded, because three bright horizontal bands spanning the full width of the sensor appeared during that exposure and its integrated profile is 0.55 of that of the other four frames. Rounding 514.9 to 515~nm in the Mie computation moves the computed maxima by less than $0.01^\circ$; repeating the complete analysis, including the calibration, with the nominal 520~nm of the part designation changes no value of $R$ by more than 0.013 and no irradiated minus non-irradiated difference by more than 0.009, with all signs unchanged. The profiles are sampled at the pixel pitch of the screen image, ${\approx}0.025^\circ$ per point, and span $\theta$ from 0 to between $19^\circ$ and $23^\circ$; the range used for analysis is set in Section~\ref{sec2.8}. Representative scattering images for every condition are shown in Supplementary Fig.~S2.

\subsection{Retention metric}\label{sec2.8}

Particle retention in the probe volume was quantified by the integrated retention
\begin{equation}
    R(t) = \frac{L^2 \int_{\theta_{\min}}^{\theta_{\max}} I(\theta,t)\,\mathrm{d}\theta}{L_{\mathrm{ref}}^2 \int_{\theta_{\min}}^{\theta_{\max}} I_{\mathrm{ref}}(\theta)\,\mathrm{d}\theta},
    \label{eq:retention}
\end{equation}
where $I_{\mathrm{ref}}$ is the reference profile of the same concentration (Section~\ref{sec2.6}), $L$ and $L_{\mathrm{ref}}$ are the fitted distances of the two measurements (Section~\ref{sec2.7}), and the integrals are evaluated by trapezoidal integration. The detector is saturated by the transmitted probe beam over a small central disc whose extent was measured directly for each of the 56 acquisitions that enter the retention analysis, that is, the 48 treated measurements of Table~\ref{tbl1}, the four of the 2.5\,fM series and the four reference measurements: it never extends beyond 1.9$^\circ$, with a median of 1.3$^\circ$, and in the three most depleted samples (2\,fM, acoustic agitation, 1, 2, and 3\,h) the azimuthally averaged profile does not reach the 8-bit ceiling at all. In every case the saturated region ends well below 3.5$^\circ$ (Supplementary Fig.~S3). The region immediately beyond is dominated by the steep forward diffraction lobe, where small errors in the pattern center produce large intensity errors. We therefore set $\theta_{\min}=3.5^\circ$. At the upper end the signal has fallen to 7\% of its value at $\theta_{\min}$ by $18^\circ$ and to 4\% by $21^\circ$, and the outer part of the pattern increasingly falls beyond the edge of the screen: at $18^\circ$, a median of 88\% of the azimuthal directions still lie on the screen. We therefore set $\theta_{\max}=18^\circ$, which excludes 3.5\% of the integral out to $21^\circ$. Each profile, sampled at $\approx 0.025^\circ$ per point ($\approx$570 points over this range), is interpolated onto a uniform 300-point grid and integrated by the trapezoidal rule. The dependence of $R$ on the integration limits and on particle number is analyzed in Section~\ref{sec3.2}.

The uncertainty of $R$ was quantified from the data itself rather than estimated, and separates into three terms of different character.

\emph{(i) Acquisition noise.} The five images of a single acquisition, taken within 30~s of each other at fixed camera settings, give integrated profiles that scatter by 1.3\% (mean standard deviation over the 56 acquisitions). This term measures the camera and the probe beam, not the experiment, and is the smallest of the three.

\emph{(ii) Aliquot-to-aliquot scatter.} This is the dominant term, and it is obtained directly from the control arm. At 3\,fM, magnetic stirring suppresses settling almost completely and irradiation has no resolvable effect under stirring (Section~\ref{sec3.1}); the eight independent aliquots of the stirring and stirring${}+{}$fs conditions are therefore all expected to return the same value of $R$. Their standard deviation is 0.022. Part of this spread is residual slow settling rather than scatter, so $\sigma=0.022$ is a conservative estimate of the uncertainty of a single $R$ value, and ${\approx}0.031$ for the difference between two independent values.

\emph{(iii) Reference scale.} All conditions of one concentration share a single reference (Section~\ref{sec2.6}), which enters every $R$ of that concentration as a common divisor; an error in the reference therefore rescales $R$ and $\Delta R$ of that concentration by a common factor but cannot alter the sign of $\Delta R$.

Because (iii) is a scale factor and (i) and (ii) are random, the primary test used in this work is distribution-free: the sign of $\Delta R$ across the twelve paired irradiated/non-irradiated comparisons (three concentrations $\times$ four treatment times). A sign test on these twelve signs requires no error model and is insensitive to the integration limits and to the angular calibration. Parametric differences are quoted alongside it, using $\sigma=0.022$.

\subsection{Temperature-dependence measurements}\label{sec2.9}

To test whether the scattering readout depends on temperature independently of sedimentation, suspensions of each concentration were prepared in deionized water pre-equilibrated at 25, 30, 35, 40, 45, and 50~$^\circ$C on a calibrated hot plate. Once the hot plate had reached the set temperature, the water was left on it for a few more minutes so that the liquid could equilibrate; the liquid temperature itself was not measured, so the quoted values are hot-plate set temperatures and the actual liquid temperature may deviate from them. PS8 stock was then added to reach the target concentration, the sample was vortexed, and scattering was acquired immediately as described above. Because this experiment is used only to show that the scattering readout does not depend on temperature, a systematic offset between the set and liquid temperatures would not affect the conclusion.

\subsection{Scaling analysis}\label{sec2.10}

Candidate mechanisms were evaluated by order-of-magnitude estimates using the particle properties given above and the properties of water at 22~$^\circ$C, representative of the laboratory temperature: dynamic viscosity $\mu = 0.955$~mPa\,s, kinematic viscosity $\nu = 9.6\times10^{-7}$~m$^2$/s, thermal diffusivity $\kappa = 1.43\times10^{-7}$~m$^2$/s, and thermal expansion coefficient $\beta = 2.3\times10^{-4}$~K$^{-1}$. The Mie scattering cross section of the particles at 515~nm was calculated for a PS refractive index of 1.599 in water ($n=1.335$).

%% ---------------------------------------------------------------- 3 RESULTS
\section{Results and Discussion}\label{sec3}

\subsection{Retention under the four processing conditions}\label{sec3.1}

Figure~\ref{fig3} shows representative scattering profiles at 3\,fM, comparing the reference with the profile after 3\,h of treatment. Under acoustic agitation alone, the intensity decreases across the full angular range, and $R(3\,\mathrm{h})=0.53$. With simultaneous irradiation, the decrease is markedly smaller ($R=0.70$). Under magnetic stirring, both profiles remain close to the reference ($R=0.93$ without and $0.94$ with irradiation). In all cases, the profiles decrease in amplitude while the angular positions of the maxima and minima are unchanged (Section~\ref{sec3.3}).

\begin{figure*}[t]
  \centering
  \includegraphics[width=0.7\textwidth]{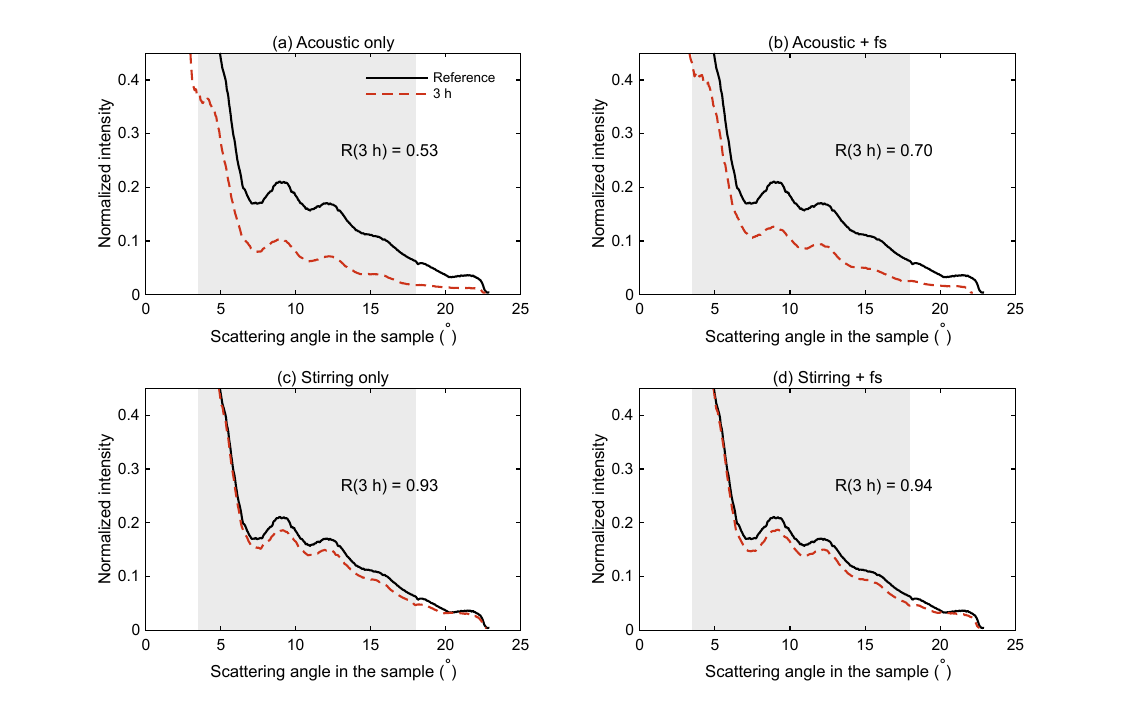}
  \caption{Representative angle-resolved Mie scattering profiles for 3\,fM PS microsphere suspensions. Each panel compares the reference measurement (solid line) with the profile after 3\,h of treatment (dashed line) for (a) acoustic agitation only, (b) acoustic agitation with fs irradiation, (c) magnetic stirring only, and (d) magnetic stirring with fs irradiation. The angular positions of the scattering maxima are preserved under all conditions. Intensities are given as a fraction of the 8-bit full scale; the shaded band marks the integration window, $\theta = 3.5$--$18^\circ$.}
\end{figure*}

The time dependence of $R$ for all conditions is shown in Fig.~\ref{fig4} and Table~\ref{tbl1}. Without agitation, particles starting at the liquid surface would settle below the probe beam, located at mid-height, within ${\approx}50$~min, and the whole ${\approx}11$~mm column would be cleared in ${\approx}1.7$~h; this is consistent with the rapid decay of unagitated controls. Under acoustic agitation alone, retention decreases rapidly during the first hour and then more slowly, reaching 0.53, 0.51, and 0.49 at 3\,h for 3, 2, and 1\,fM, respectively.

\begin{figure*}[t]
  \centering
  \includegraphics[width=0.5\textwidth]{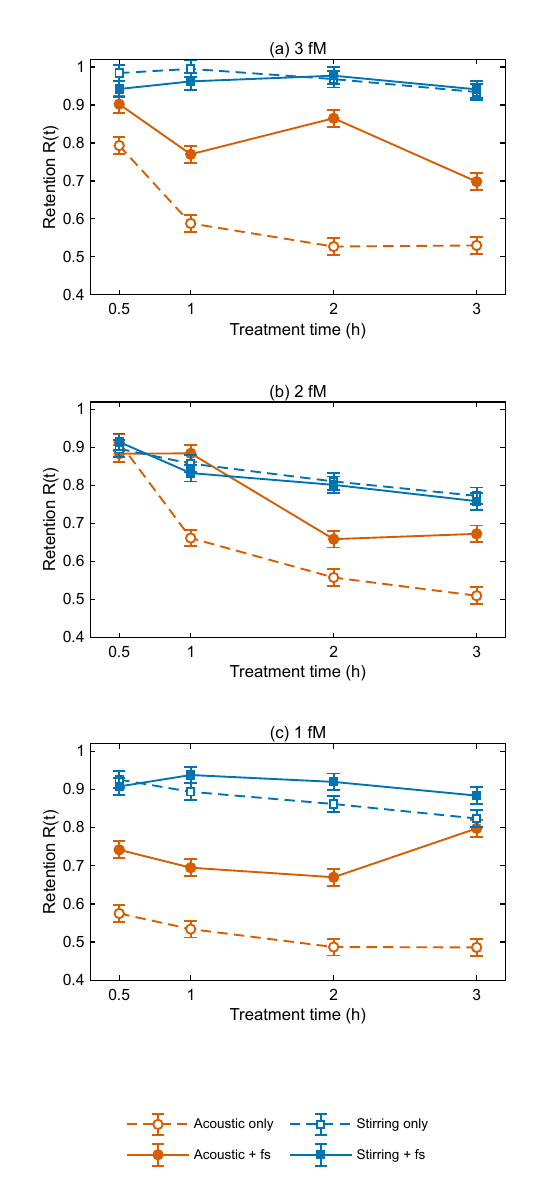}
  \caption{Integrated scattering retention $R(t)$ as a function of treatment time for (a) 3\,fM, (b) 2\,fM, and (c) 1\,fM PS microsphere suspensions under acoustic agitation only, acoustic agitation with fs irradiation, magnetic stirring only, and magnetic stirring with fs irradiation. Each point is an independent aliquot; error bars indicate the measured aliquot-to-aliquot standard deviation of $0.022$ (Section~\ref{sec2.8}). Lines are guides to the eye.}
\end{figure*}

\begin{table}[pos=ht]
\caption{Integrated scattering retention $R(t)$ for all conditions ($\theta = 3.5$--$18^\circ$). Each value is an independent aliquot; the uncertainty of a single value is $\sigma = 0.022$ (Section~\ref{sec2.8}). Within a concentration, all four conditions are normalized to one and the same reference measurement, so that an error in the reference rescales $\Delta R$ by a common factor. $\overline{\Delta R}$ is the mean of the irradiated minus non-irradiated difference over the four treatment times.}\label{tbl1}
\centering\small
\setlength{\tabcolsep}{6pt}
\renewcommand{\arraystretch}{1.12}
\begin{tabular}{@{}llccccc@{}}
\toprule
Conc. & Condition & $R$(0.5\,h) & $R$(1\,h) & $R$(2\,h) & $R$(3\,h) & $\overline{\Delta R}$ \\
\midrule
3\,fM & Acoustic & 0.793 & 0.588 & 0.527 & 0.530 &  \\
 & Acoustic + fs & 0.902 & 0.770 & 0.865 & 0.698 & $+0.199$ \\
 & Stirring & 0.984 & 0.995 & 0.968 & 0.934 &  \\
 & Stirring + fs & 0.942 & 0.962 & 0.977 & 0.941 & $-0.014$ \\
\midrule
2\,fM & Acoustic & 0.914 & 0.662 & 0.558 & 0.510 &  \\
 & Acoustic + fs & 0.884 & 0.885 & 0.659 & 0.673 & $+0.114$ \\
 & Stirring & 0.897 & 0.858 & 0.811 & 0.773 &  \\
 & Stirring + fs & 0.915 & 0.833 & 0.802 & 0.759 & $-0.007$ \\
\midrule
1\,fM & Acoustic & 0.575 & 0.534 & 0.487 & 0.486 &  \\
 & Acoustic + fs & 0.742 & 0.695 & 0.670 & 0.798 & $+0.206$ \\
 & Stirring & 0.926 & 0.894 & 0.862 & 0.824 &  \\
 & Stirring + fs & 0.908 & 0.938 & 0.920 & 0.884 & $+0.036$ \\
\bottomrule
\end{tabular}
\end{table}

The central result is the comparison between the irradiated and non-irradiated condition at matched agitation, concentration, and treatment time. Under acoustic agitation, eleven of these twelve paired comparisons are positive (Table~\ref{tbl1}): irradiation increases retention at every concentration and, except at the shortest treatment time at 2\,fM, at every treatment time. Taken as twelve independent signs, this has a two-sided probability of $6\times10^{-3}$ under the null hypothesis of no effect, without any assumption about the size of the experimental uncertainty. The magnitude of the increase averages $+0.17$ overall, and $+0.20$, $+0.11$, and $+0.21$ at 3, 2, and 1\,fM, with a largest single value of $+0.34$ (3\,fM, 2\,h); the eleven positive differences all exceed $2\sigma$ of a difference (0.062), the smallest being $+0.10$ (2\,fM, 2\,h). The single negative difference, $-0.03$ (2\,fM, 0.5\,h), lies within the aliquot-to-aliquot scatter. The 1\,fM comparison under acoustic agitation rests on a normalization across a reassembled setup (Section~\ref{sec2.7}); without it, eight of the twelve comparisons are positive, and seven of the eight at 2 and 3\,fM.

The matched stirring arm behaves as a control and shows no such effect. There, six of the twelve differences are positive and six negative, and the mean differences are $-0.01$, $-0.01$, and $+0.04$ at 3, 2, and 1\,fM, an order of magnitude smaller than under acoustic agitation. None of the twelve stirring differences exceeds $2\sigma$, against eleven of the twelve in the acoustic arm. The contrast between the two arms is the strongest evidence that the increase under acoustic agitation originates in the irradiation and not in sample handling, batch-to-batch variation, or the measurement chain, all of which are common to the two arms.

A fourth concentration supports the same picture. An additional series at 2.5\,fM under acoustic agitation with irradiation, measured against its own reference in the same way, gives $R = 0.867$, 0.765, 0.561, and 0.611 at 0.5, 1, 2, and 3\,h (Supplementary Table~S1). No matched non-irradiated control was run at this concentration, so it cannot contribute a paired comparison and it is not included in Table~\ref{tbl1}; we report it because it is an independent realization of the irradiated acoustic condition, prepared and measured on different days from the others. Taken together, the irradiated acoustic condition was realized four times, at 1, 2, 2.5, and 3\,fM, giving $R(3\,\mathrm{h})$ between 0.61 and 0.80, while the non-irradiated acoustic condition was realized three times, giving 0.49 to 0.53. The two sets of independent realizations do not overlap.

The effect of irradiation therefore depends mainly on the strength of the mechanical agitation. Stirring at 500\,rpm already keeps most particles in suspension ($R > 0.75$ at every concentration and time), leaving little room for an additional contribution; the weaker and less homogeneous acoustic agitation leaves a larger fraction of particles to settle, and the additional fluid motion provided by irradiation (Section~\ref{sec3.5}) produces a large relative gain. This is consistent with the observation that optical perturbations are most effective when baseline mixing is insufficient \citep{Seo2015}. 

Two further features of the data are noteworthy. First, the individual series are not monotonic in treatment time: because each point is an independent aliquot rather than a later observation of the same sample (Section~\ref{sec2.6}), aliquot-to-aliquot scatter produces apparent increases with time, the largest being $+0.13$ at 1\,fM between 2 and 3\,h under acoustic agitation with irradiation. This is why the analysis above rests on the twelve paired comparisons rather than on the shape of any single curve. Second, the magnitude of the effect does not vary monotonically with concentration ($+0.21$, $+0.11$ and $+0.20$ at 1, 2 and 3\,fM); we do not interpret this ordering.

\subsection{Robustness of the retention metric}\label{sec3.2}

\textit{Relation between $R$ and particle number.} For identical particles in the single-scattering limit, the intensity scattered into any angle is proportional to the number $N$ of particles in the probe volume, so that $R = N/N_0$. At the present concentrations, attenuation of the probe beam within the cell is not negligible. From Mie theory, the scattering efficiency of the particles at 515~nm is $Q_{\mathrm{sca}}=2.03$ ($C_{\mathrm{sca}}=1.02\times10^{-10}$~m$^2$), of which 62\% is scattered beyond $3.5^\circ$. Over the optical path of $L=10$~mm, the optical depth for scattering beyond $3.5^\circ$ is $\tau_0 = n_0 C_{\mathrm{sca},>3.5^\circ}L = 0.38$, 0.76, and 1.15 at 1, 2, and 3\,fM. Attenuation of the incident and scattered light by the other particles reduces the signal of a dense sample more than that of a dilute one, so that, to first order, $R = \nu\,\exp[\tau_0(1-\nu)]$ with $\nu=N/N_0$. $R$ is thus a monotonic function of $\nu$, and $R\geq\nu$: the metric underestimates particle loss but preserves the ordering of conditions. Because $R/\nu$ decreases monotonically with $\nu$, differences in particle number between conditions are at least as large, in relative terms, as the differences in $R$. For example, at 3\,fM and 3\,h, $R=0.53$ and $0.70$ correspond to $\nu \approx 0.22$ and $0.32$, i.e., irradiation increases the number of particles retained in the probe volume by a factor of ${\approx}1.5$. The corresponding values for all conditions are listed in Supplementary Table~S2. The same attenuation is qualitatively consistent with the sublinear increase of the reference intensities with concentration, which rise approximately as $1:1.12:1.18$ for $1:2:3$\,fM (the three reference measurements were recorded on different days, so this comparison is indicative rather than quantitative). All conclusions drawn in this work rest on comparisons of $R$ between conditions and are therefore conservative.

\textit{Integration limits.} To test the dependence of the result on the angular range, the whole analysis was repeated for the integration windows [3.0$^\circ$, 18$^\circ$], [5.0$^\circ$, 18$^\circ$], [3.5$^\circ$, 14$^\circ$], [3.5$^\circ$, 20$^\circ$], and [6.0$^\circ$, 16$^\circ$], which differ substantially in both limits and in the number of Mie orders they contain (Supplementary Table~S3). The mean irradiated minus non-irradiated difference under acoustic agitation, averaged over the three concentrations, is $+0.173$ for the window used here and ranges only from $+0.159$ to $+0.205$ across the five alternatives. Eleven of the twelve paired comparisons are positive in every window, and all twelve in [6.0$^\circ$, 16$^\circ$]. The result is therefore not an artifact of the chosen angular range.

\textit{Temperature.} Irradiation deposits heat in the sample (Section~\ref{sec3.4}), so a temperature dependence of the measurement itself would be a confound. The scattering of freshly prepared suspensions was therefore measured between 25 and 50\,$^\circ$C (Fig.~\ref{fig5}). At 3\,fM the integrated intensity shows no trend with temperature, scattering between $-7\%$ and $-1\%$ of its value at 25\,$^\circ$C. At 2\,fM every point above 25\,$^\circ$C lies $10$--$16\%$ higher than the 25\,$^\circ$C value but without a trend among themselves. At 1\,fM the behavior is different: after a dip of up to 5\% at 30 and 35\,$^\circ$C the intensity rises monotonically, reaching $+15\%$ at 50\,$^\circ$C. This rise is systematic and larger than the measurement scatter; its origin has not been established.

Whatever its origin, the temperature response cannot generate the reported differences, because the stirring arm bounds it directly. The stirring${}+{}$fs samples receive the same irradiation, and therefore the same heating, as the acoustic${}+{}$fs samples, yet show a mean difference of $+0.01$ against their non-irradiated counterparts. If elevated temperature at the moment of measurement inflated $R$, it would do so equally in both arms, which is not observed. The largest stirring-arm mean difference is at 1\,fM ($+0.04$), the concentration with the largest temperature sensitivity, which sets a conservative upper bound of $\approx$0.04 on any thermal contribution to the acoustic-arm differences at that concentration, $+0.16$ to $+0.31$.

\textit{Evaporation.} Because the irradiated samples are heated for up to 3\,h, loss of water could in principle concentrate the suspension and raise $R$ without any change in the behavior of the particles. The cuvettes were sealed with Parafilm throughout treatment, and no change of the liquid level was observed, but the effect is also bounded quantitatively. To raise $R$ by $\approx$0.2 on a base of $\approx$0.5 would require the particle concentration to increase by $\approx$40\%, i.e.\ the loss of $\approx$0.31\,mL of the 1.1\,mL charge. That corresponds to $\approx$0.71\,kJ of latent heat, 19\% of the entire 3.8\,kJ absorbed by the sample over 3\,h, and to a fall of the liquid column from 11 to $\approx$7.9\,mm, which would have been readily visible. Neither was observed. Moreover, the stirring${}+{}$fs samples are heated identically and would concentrate identically, yet show no increase. Evaporation can therefore be excluded as the origin of the effect.

\begin{figure*}[t]
  \centering
  \includegraphics[width=0.5\textwidth]{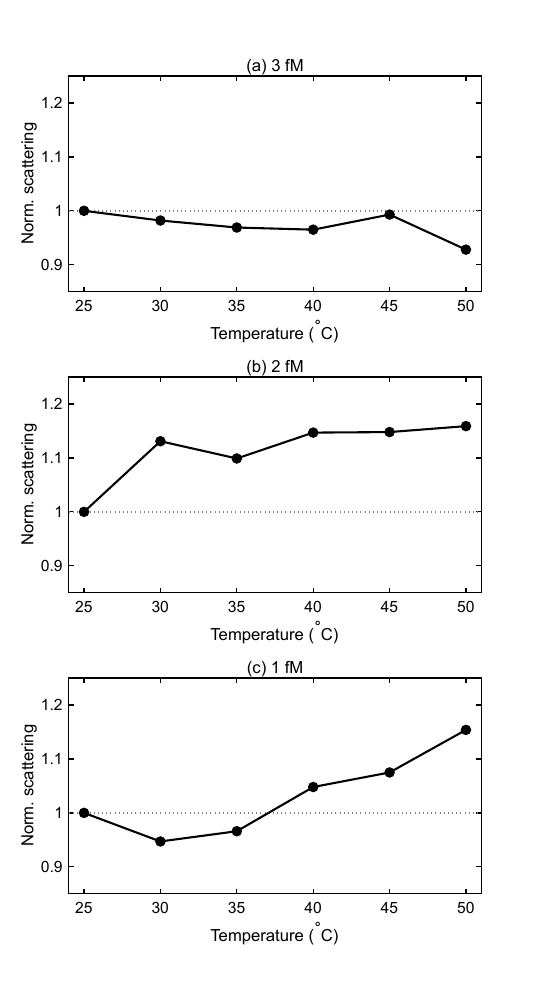}
  \caption{Temperature dependence of the integrated scattering intensity of freshly prepared PS microsphere suspensions at (a) 3\,fM, (b) 2\,fM, and (c) 1\,fM, prepared in water pre-equilibrated at hot-plate set temperatures between 25\,$^\circ$C and 50\,$^\circ$C and normalized to the value at 25\,$^\circ$C. The response shows no trend at 3\,fM ($-7\%$ to $-1\%$) or at 2\,fM (all points $10$--$16\%$ above the 25\,$^\circ$C value, with no trend among themselves); at 1\,fM the intensity rises monotonically from 30 to 50\,$^\circ$C, reaching $+15\%$.}
\end{figure*}

\subsection{Preservation of particle size}\label{sec3.3}

The angular positions of the Mie maxima are determined by the ratio of particle size to wavelength and by the relative refractive index, and scale as $1/d$ for small changes in particle diameter $d$ \citep{Hulst1981,Genc2023}. Because the angular axis of each profile is itself calibrated against the Mie pattern (Section~\ref{sec2.7}), the absolute positions cannot by themselves test for a change in $d$: a uniform size change would simply be absorbed into the fitted $L$. What can be tested, and what is physically meaningful, is whether $L$ drifts systematically with treatment time within a series.

The fitted $L$ conflates two effects: changes in the cuvette-to-screen geometry, which vary from measurement to measurement because the cell is removed and replaced, and changes in particle size, which under progressive fragmentation, ablation, or aggregation would accumulate with exposure. Within the twelve series of Table~\ref{tbl1}, 48 calibrations in all, $L$ scatters by 0.6\%, and its regression against treatment time gives a slope of $+0.20 \pm 0.08$\% per hour ($t = 2.4$; Supplementary Table~S4). The drift is not specific to irradiation: the series slopes average $+0.36$\% per hour with irradiation and $+0.04$\% per hour without, a difference that is not significant ($t = 1.7$), and a slow change of the geometry over a measurement day would produce the same signature. Taking the 95\% bound on the pooled slope over the 3~h exposure, and attributing all of it to the particles, any systematic change in particle diameter is smaller than 1.1\%. Fragmentation, ablation, or aggregation at a level that would alter the scattering of the population can therefore be excluded, and the measured decrease in $R$ reflects the number of particles in the probe volume rather than a change in the particles themselves. The peak fluence at the focus (${\approx}9.4$~J/cm$^2$) is several times the reported single-pulse femtosecond ablation threshold of polystyrene, $3.2 \pm 0.3$~J/cm$^2$ for intrinsic material and $0.40 \pm 0.07$~J/cm$^2$ for defect-mediated ablation \citep{Naghilou2015}; those values were obtained on high-impact polystyrene with sub-30~fs pulses at 790~nm, so the comparison is an order-of-magnitude one, but it indicates that a particle reaching the focus is likely to be modified. This region, however, occupies only ${\sim}10^{-9}$ of the sample volume (Section~\ref{sec2.3}), and the unchanged peak positions show that any such modification affects a negligible fraction of the particle population. This distinguishes the present effect from laser stabilization of nanoparticle suspensions, which relies on fragmentation of particles or clusters \citep{Seo2015,Dai2017}.

\subsection{Heat deposition}\label{sec3.4}

Although water is often described as transparent in the near infrared, its absorption at 1030~nm is not negligible on a centimeter path. Tabulated values give a linear absorption coefficient of $\alpha \approx 0.22$~cm$^{-1}$ at this wavelength \citep{HaleQuerry1973,Kou1993}. Over the 1~cm optical path in the cell, a fraction $1-e^{-\alpha L} \approx 20\%$ of the incident 1.8~W, i.e.\ ${\approx}0.35$~W, is therefore deposited in the water along the beam, as a line heat source spanning the full width of the cell. Near the focus, nonlinear absorption adds to this amount \citep{Vogel2005}.

This is a large thermal load for a 1.1~mL sample, whose heat capacity is only ${\approx}4.6$~J/K: in the absence of heat loss the liquid would warm at ${\approx}4.5$~K/min. The steady state is set by heat loss through the cuvette walls, the sealing film, and conduction into the holder and translation stage. Taking only natural convection and radiation from the ${\approx}5$~cm$^2$ of exposed cuvette surface ($h \approx 10$--$15$~W\,m$^{-2}$K$^{-1}$) gives a steady-state rise of several tens of kelvin, approached within tens of minutes; conduction into the holder, which was not quantified, reduces this. The temperature rise was not measured \emph{in situ}, and the estimate above should be read as an upper bound rather than a value. It is consistent with the two qualitative observations reported in Section~\ref{sec2.3}: the samples were noticeably warm at the end of treatment, and gas bubbles appeared progressively throughout the liquid and on the walls, as expected from the release of dissolved air in warming water rather than from cavitation at the focus. It also explains the choice of 25--50~$^\circ$C as the range for the temperature control experiment of Section~\ref{sec3.2}. For the present argument, the relevant point is not the magnitude of the temperature rise but the fluid motion that even a small fraction of it must drive, as discussed next.

\subsection{Mechanism of stabilization}\label{sec3.5}

A particle remains in the probe volume only if the local fluid velocity has an upward component comparable to or larger than its Stokes settling velocity. For the present particles,
\begin{equation}
    v_s = \frac{2\,\Delta\rho\, g\, a^2}{9\mu} \approx 1.8~\mu\text{m/s} \;(6.6~\text{mm/h}),
\end{equation}
and the P\'eclet number $v_s a/D \approx 130$, so Brownian motion is irrelevant. Any mechanism proposed for the stabilization must therefore produce sustained fluid or particle velocities of at least a few $\mu$m/s throughout a substantial part of the sample. We evaluate the candidate mechanisms against this criterion.

\textit{Optical forces.} The buoyant weight of a particle is only 0.13~pN, and optical forces from a watt-level beam can exceed it. However, the scattering force acts along the horizontal beam axis and pushes particles toward the rear wall rather than upward, and the transverse gradient force can hold a particle against gravity only near the focus, where the intensity gradient is large. The converging and diverging beam cone (half-angle ${\approx}4^\circ$ in water) occupies only ${\approx}0.1\%$ of the sample volume, and the focal region, where the intensity gradient is large, only ${\sim}10^{-9}$ of it. At any time, therefore, only a negligible fraction of the particles is subject to an appreciable optical force, which cannot account for retention changes of 0.1--0.3 across the whole particle population.

\textit{Optical Kerr effect and transient photochemistry.} Multiphoton absorption at the focus generates solvated electrons and reactive species, and the Kerr effect creates transient refractive index gradients \citep{Danilov2025,Betka2025,Righini1993}. These processes are confined to the focal region and decay within picoseconds to nanoseconds \citep{Vogel2005}. Their direct momentum transfer to the fluid is negligible; their effect on the fluid enters only through the heat and bubbles they leave behind, which are considered below.

\textit{Thermal convection.} A horizontal temperature difference drives buoyancy-driven convection without threshold \citep{Riviere2016}. In the viscous regime appropriate for the present cell (width 10~mm, liquid height ${\approx}11$~mm), the characteristic velocity over a length scale $L$ is
\begin{equation}
    u \sim \frac{g\beta\,\Delta T\,L^2}{\nu}.
\end{equation}
For $L=5$~mm (half the cell width), a temperature difference of only $\Delta T=0.01$~K gives $u\approx0.6$~mm/s, more than two orders of magnitude above $v_s$; even allowing for a large overestimate by this scaling in a confined geometry, temperature differences of millikelvins suffice to produce convective velocities exceeding the settling velocity. The 0.35~W absorbed along the beam (Section~\ref{sec3.4}) produces horizontal temperature differences many orders of magnitude larger than 0.01~K. Laser-driven convection is therefore therefore expected under the present conditions, and it is the only mechanism of sufficient magnitude to act on the whole particle population. Similar optothermal flows transport microparticles at tens of $\mu$m/s in thin cells at milliwatt to watt power levels \citep{Li2025,RamirezRamirez2023,LiuPoon2010}.

\textit{Bubbles.} Bubbles appeared only after some time of irradiation, were distributed throughout the liquid and on the cuvette walls rather than confined to the focus, remained largely stationary, and increased in number with exposure time (Section~\ref{sec2.3}). This behavior is characteristic of the release of dissolved air as the water warms, rather than of transient cavitation at the focus \citep{Vogel2005,Sabzeghabae2021}, and is itself an indication of heating of the bulk liquid. Because the bubbles were largely stationary, they are unlikely to be the main source of fluid motion, although their nucleation and growth may contribute locally \citep{Lauterborn2010}.

Taken together, the estimates show that fs irradiation stabilizes the suspension by driving fluid motion in the medium, predominantly thermal convection with a possible contribution from bubbles, rather than by acting directly on the particles. The role of the fs pulses is to deposit energy in the water, both linearly along the beam and nonlinearly at the focus, and to generate bubbles, without altering the particles (Section~\ref{sec3.3}). This picture explains the main observations: the irradiation acts as a supplementary, non-contact agitation source whose contribution is large when the mechanical agitation is weak (acoustic agitation) and small when it already maintains most particles in suspension (stirring at 3\,fM). The temperature experiment excludes a thermal artifact in the readout, whereas heating is the most likely driver of the stabilizing flow itself.

\textit{Limitations.} Access to the fs laser system ended after the measurements reported here. Three limitations follow from this. First, no treated condition was repeated at the same concentration and treatment time, so there is no direct replicate of an individual value; the uncertainty is instead measured from the control arm (Section~\ref{sec2.8}), and the central result is stated as a sign test over twelve paired comparisons so that it does not depend on any single value. The 1\,fM acoustic series was recorded in a reassembled scattering setup, and its comparison with the irradiated series depends on the reference to which it is normalized. Second, the mechanism is identified by scaling analysis rather than by measurement: the flow field was not visualized and the temperature rise was not measured \emph{in situ}, so the partition between thermal convection and bubble-driven flow remains open, although both are laser-driven fluid motion and the conclusion does not depend on how they are apportioned. Third, and most importantly, no control was run with continuous-wave or low-repetition-rate irradiation at equal average power. On the mechanism proposed here the femtosecond character of the pulses is not expected to be essential, since the heat is deposited by linear absorption of the average power; a continuous-wave source of equal absorbed power should therefore produce a comparable effect. This prediction remains to be tested. High-speed imaging or particle tracking of the flow, time-resolved shadowgraphy, \textit{in situ} thermometry, a bubble-suppression experiment, spatially resolved concentration measurements, and irradiation with a continuous-wave or low-repetition-rate source at equal average power are the natural next steps.

%% ---------------------------------------------------------------- 4 CONCLUSION
\section{Conclusion}\label{sec4}

Femtosecond laser irradiation at 1030~nm retards the sedimentation of 8~$\mu$m PS microspheres in deionized water. Under acoustic agitation, irradiation raised retention in eleven of twelve paired comparisons spanning three concentrations and four treatment times, by $0.17$ on average, a result that follows from the signs alone ($p = 6\times10^{-3}$) and requires no assumption about the experimental uncertainty; the twelfth comparison lies within the measured scatter. At 3\,fM and 3\,h this corresponds to an increase of ${\approx}1.5$ in the number of particles retained in the probe volume. Under magnetic stirring, which already maintains most particles in suspension and which receives the same irradiation and the same heating, the mean difference is $0.01$ and the signs are evenly split; it is this matched control that attributes the effect to the irradiation. Irradiation never reduced retention beyond the measurement uncertainty.

The integrated Mie scattering metric $R(t)$ provides a non-invasive, conservative measure of retention whose conclusions are unchanged across five alternative integration windows, and the absence of any systematic drift of the Mie peak positions with treatment time bounds changes in particle size at about 1\% over 3~h. A scaling analysis shows that optical forces, the Kerr effect, and transient photochemistry cannot account for the observed retention, whereas the power absorbed in the water drives convective flows that exceed the Stokes settling velocity by orders of magnitude. The stabilization is therefore attributed to laser-driven fluid motion, predominantly thermal convection, with a possible contribution from laser-generated bubbles.

Femtosecond irradiation thus acts as a non-contact, chemically non-invasive supplementary agitation source that preserves particle identity and is most effective where mechanical mixing is weak. Beyond the present model system, the combination of this approach with angle-resolved scattering may be useful for keeping and monitoring suspensions of microorganisms, cells, or environmental microparticles in a sealed optical cell \citep{Felter1969,Fradette2022,Genc2023}. Direct flow visualization, \emph{in situ} thermometry, and comparison with continuous-wave heating will allow the contributions of convection and bubbles to be quantified.

\section*{CRediT authorship contribution statement}
\textbf{Hafsa Moazzamia:} Investigation, Methodology, Formal analysis, Writing -- original draft. 
\textbf{Irem Hazal Tan:} Investigation, Software, Data curation, Methodology. \textbf{Alpan Bek:} Conceptualization, Resources, Writing -- review \& editing. 
\textbf{Sinan Genc:} Conceptualization, Supervision, Methodology, Formal analysis, Writing -- original draft, Writing -- review \& editing.

\section*{Declaration of competing interest}
The authors declare no conflicts of interest.

\section*{Declaration of generative AI and AI-assisted technologies in the writing process}
During the preparation of this work, the authors used Claude (Anthropic) solely to improve the language, readability, and flow of the manuscript text. The tool was not used to generate scientific content, data, results, figures, or interpretations. After using this tool, the authors reviewed and edited the content as needed and take full responsibility for the content of the publication.

\section*{Data and code availability}
  All raw camera frames, the analysis code and the figure scripts are openly available at Zenodo, \url{https://doi.org/10.5281/zenodo.23285962}, the code under the MIT license and the data under CC~BY~4.0.

\section*{Acknowledgment}
The authors thank Dr.\ Arian Goodarzi and Asst.\ Prof.\ Ihor Pavlov for providing access to the femtosecond laser system, and Mohammad Moharrami for his assistance with the irradiation setup.

\section*{Funding}
This work was supported by the Scientific and Technological Research Council of T\"urkiye (T\"UB\.ITAK) under project no.\ 122E144.

\bibliographystyle{unsrtnat}
\bibliography{sample}

\end{document}